\documentclass[jnr]{iosart2x}

\usepackage[utf8]{inputenc}

\usepackage{subcaption}
\usepackage{listings}
\definecolor{darkred}{rgb}{0.6,0,0}
\definecolor{darkblue}{rgb}{0,0,.6}
\definecolor{darkgreen}{rgb}{0,.6,0}
\definecolor{operator}{rgb}{0.6,0.3,1.0}
\usepackage{xspace}
\newcommand{\variable}{\texttt{Variable}\xspace}
\newcommand{\dataarray}{\texttt{DataArray}\xspace}
\newcommand{\dataset}{\texttt{Dataset}\xspace}
\newcommand{\mantid}{Mantid\xspace}
\newcommand{\pandas}{pandas\xspace}
\newcommand{\xarray}{xarray\xspace}
\newcommand{\scipp}{\texttt{scipp}\xspace}

\newcommand{\xtensor}{xtensor\xspace}

\newcommand{\numpy}{NumPy\xspace}
\newcommand{\cpp}{C++\xspace}
\newcommand{\python}{Python\xspace}

\newcommand{\conda}{\texttt{conda}\xspace}
\newcommand{\jupyter}{Jupyter\xspace}

\newcommand{\matplotlib}{Matplotlib\xspace}
\newcommand{\nexus}{NeXus\xspace}

\usepackage[many]{tcolorbox}
\tcbuselibrary{listings}
\definecolor{light-gray}{gray}{0.95}

\newcounter{ipythcntr}

\newtcblisting{ipythonnb}[1][\theipythcntr]{
  arc=0.5pt,
  width=\linewidth,
  enhanced,
  boxrule=0.4pt,
  boxsep=-3mm,
  colback=light-gray,
  listing only,
  top=0pt,
  bottom=0pt,
  listing options={
  frame=,
  basicstyle=\scriptsize\ttfamily,
  },
}
\pubyear{0000}
\volume{0}
\firstpage{1}
\lastpage{1}

\begin{document}
\makeatletter
\let\put@numberlines@box\relax
\makeatother

\begin{frontmatter}

%\pretitle{}
\title{Scipp: Scientific data handling with labeled multi-dimensional arrays for C++ and Python}
\runningtitle{Proceedings of ICANS-XXIII}
%\subtitle{}

% For one author:
%\author{\inits{N.}\fnms{Name1} \snm{Surname1}\ead[label=e1]{first@somewhere.com}}
%\address{Department first, \orgname{University or Company name},
%Abbreviate US states, \cny{Country}\printead[presep={\\}]{e1}}
%\runningauthor{N. Surname1}

% Two or more authors:
\author[A]{\inits{S.}\fnms{Simon} \snm{Heybrock}\ead[label=e1]{simon.heybrock@ess.eu}%
\thanks{Corresponding author. \printead{e1}.}},
\author[B]{\inits{O.}\fnms{Owen} \snm{Arnold}\ead[label=e2]{owen.arnold@tessella.com}},
\author[A]{\inits{I.}\fnms{Igor} \snm{Gudich}\ead[label=e3]{igor.gudich@ess.eu}},
\author[C]{\inits{D.}\fnms{Daniel} \snm{Nixon}\ead[label=e4]{daniel.nixon@stfc.ac.uk}}
and
\author[A]{\inits{N.}\fnms{Neil} \snm{Vaytet}\ead[label=e5]{neil.vaytet@ess.eu}}
\runningauthor{S. Heybrock et al.}
\address[A]{Data Management and Software Centre, \orgname{European Spallation Source},
\cny{Sweden}\printead[presep={\\}]{e1,e3,e5}}
\address[B]{\orgname{Tessella},
\cny{UK}\printead[presep={\\}]{e2}}
\address[C]{\orgname{Science and Technology Facilities Council},
\cny{UK}\printead[presep={\\}]{e4}}

\begin{abstract}
  \scipp is heavily inspired by the Python library \emph{xarray}.
It enriches raw \numpy-like multi-dimensional arrays of data by adding named dimensions and associated coordinates.
Multiple arrays are combined into datasets.
On top of this, \scipp introduces
(i) implicit handling of physical units,
(ii) implicit propagation of uncertainties,
(iii) support for histograms, i.e., bin-edge coordinate axes, which exceed the data's dimension extent by one, and
(iv) support for event data.
In conjunction these features enable a more natural and more concise user experience.
The combination of named dimensions, coordinates, and units helps to drastically reduce the risk for programming errors.
The core of \scipp is written in \cpp to open opportunities for performance improvements that a \python-based solution would not allow for.
On top of the \cpp core, \scipp's \python components provide functionality for plotting and content representations, e.g., for use in \jupyter Notebooks.
While none of \scipp's concepts in isolation is novel \emph{per-se}, we are not aware of any project combining all of these aspects in a single coherent software package.
\end{abstract}

\begin{keyword}
\kwd{Software}
\kwd{Python}
\kwd{C++}
\kwd{data-processing}
\kwd{multi-dimensional}
\end{keyword}

\end{frontmatter}

%%%%%%%%%%% The article body starts:

\section{Introduction}\label{sec:introduction}

The \scipp library~\cite{scipp-rtd, scipp-github} presented here can be seen as a step that combines requirements, knowledge, and capabilities from two streams of software development that have run independently
for the past decade.

One the one hand, modern, flexible, and powerful concepts and tools have evolved (and still are evolving) in the Python ecosystem.
Prime examples are \pandas~\cite{mckinney-proc-scipy-2010}, \xarray~\cite{hoyer2017xarray}, and \jupyter~\cite{Kluyver:2016aa}.
For example, \xarray provides flexible data structures for multi-dimensional data with named dimensions, coordinates, and attributes.
This ``labeled'' way of storing data supports a good user-experience for developers and significantly reduces the risk for certain classes of bugs and errors, ensuring that scientific results are correct.

One the other hand, the Mantid framework~\cite{ARNOLD2014156} brings event-data handling for neutron-scattering experiments with support for physical units, uncertainties, and histograms.
It is used at and jointly developed by ISIS, SNS, ESS, and ILL, with additional users and contributions from other neutron sources.
One of Mantid's strong points is processing of time-of-flight neutron scattering data in event mode, i.e., every neutron detected is recorded as a timestamp and a position index.
Over the years however, new requirements arose that are inconvenient to handle in the existing framework.
Examples include:
(a) Storage of and visualization of data depending on arbitrary additional parameters, in particular scans of sample-environment parameters such as a sample temperature.
(b) Storage of and visualization of data not depending on time-of-flight (or derived dimensions), e.g., for monochromatic beamlines at reactor sources or neutron-imaging techniques.
(c) Uncertainties and units attached to all entities:
both to data and (optionally) its coordinates.
%TODO , e.g., $I(Q_x, Q_y)$ for SANS data with associated $\Delta Q_x$ and $\Delta Q_y$ on the axes.
(d) Possibility to have multiple axes along the same dimension.
(e) Exploration of opportunities for performance improvements such as more efficient data layouts, the use of single-precision, and alternative approaches to multi-threading and parallelization.

Mantid lacks the flexibility to support such requirements without resorting to workarounds and furthermore misses several other desirable features of \xarray such as a modern, simple, and coherent \python interface or good interoperability with \numpy.
With \mantid's codebase exceeding 2 million lines of code, significant changes in this direction are not feasible since they invariably become large-scale and breaking.
In turn, \xarray currently lacks support for features that are essential for neutron scattering data reduction.
In particular, the following are must-have features:

\begin{enumerate}
  \item Automatic handling of physical units.
  \item Automatic propagation of uncertainties.
  \item Support for histograms, i.e., bin-edge coordinate axes, which exceed the data's dimension extent by one.
  \item Support for event data, an array of nested random-length lists.
\end{enumerate}

We therefore chose to develop a new library, \scipp\footnote{Etymology: Scientific \cpp library $\rightarrow$ Sci++ $\rightarrow$ \scipp.}, combining above must-have features with flexible labeled data structures \emph{à la} \xarray.
We considered it infeasible to implement \emph{all} our additional requirements as contributions to the \python library \xarray. 
One of the main reasons is the central role event-data plays for us --- with \xarray being based on regular dense \numpy arrays this would have been a major challenge and would have posed unknown risks for performance.
For a more detailed discussion on the reasoning see Sec.~\ref{sec:xarray}.
\scipp's core is written in \cpp but \scipp is intended to be used mainly through its \python bindings.

The remainder of the paper is organized as follows.
Section~\ref{sec:scipp-fundamentals} gives an overview of \scipp's data structures and operations, Sec.~\ref{sec:details} covers development methodologies, the \python interface, and technical details of \scipp's architecture, and Sec.~\ref{sec:evaluation} discusses performance, limitations, and related work.
We conclude in Sec.~\ref{sec:conclusion}.

\section{Scipp fundamentals}\label{sec:scipp-fundamentals}
\subsection{Data structures and operations}\label{sec:data-structures}

The central data structures in \scipp are \variable, \dataarray, and \dataset.\footnote{The latter two are conceptually very similar to \dataarray and \dataset in the \xarray Python library.}
Variables are the building block used for data arrays and datasets.
Conceptually, a variable consists of:

\begin{description}
  \item[\texttt{values}] a multi-dimensional array of values
  \item[\texttt{variances}] a (optional) multi-dimensional array of variances for the array values
  \item[\texttt{dims}] a list of dimension labels for each axis of the array
  \item[\texttt{unit}] a physical unit of the values in the array
\end{description}

In other words, a variable represents an array-valued physical quantity.
Based on this abstraction, we implement basic element-wise operations such as addition, multiplication, or trigonometric functions.
Such operations can propagate uncertainties and handle physical units, and operations produce an error if the units of the operands are incompatible.
The dimension labels are used to implement safe slicing, safe broadcasting, and safe transposition, and operations produce an error if dimensions and/or array shapes are incompatible.

\dataarray consists of a single variable representing data, alongside dictionaries of coordinates and attributes.
Each of the coordinates and attributes are variables themselves.
The coordinate dictionary provides a mapping from a key to the corresponding coordinate.
Just like \xarray, we distinguish \emph{dimension-coordinates} with a key equal to the dimension of the coordinate, and \emph{non-dimension-coordinates} with a key that is not part of the coordinate's dimensions.
Non-dimension-coordinates are essentially auxiliary labels for an axis.
Finally, attributes are supported to store arbitrary other metadata.
Based on this, we define \emph{operations between data arrays}, one of the central concepts in \scipp:
\begin{enumerate}
  \item \emph{Compare} coordinate and label values and their units.
  \item \emph{Operate on} the data.
  \item Ignore attributes unless there is a unique way of preserving them.
\end{enumerate}
That is, a mismatch of any of the coordinates will produce an error and data is not modified.
Since data and coordinates are each internally variables, operations automatically handle physical units and propagation of uncertainties.
\dataset provides a dictionary of data arrays with \emph{aligned dimensions}.
Essentially it can be seen as a data array with multiple data entries, each identified by a name string.
Operations between datasets match data items based on their name string.

\begin{figure}
  \includegraphics[width=0.5\textwidth]{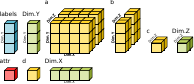}
\caption{\label{fig:show}
\scipp's auto-generated visualization of structure and content of a dataset.
The dataset contains four data arrays named ``a'', ``b'', ``c'', and ``d'' (yellow).
``a'' is a 3-D array with an associated array of variances (visualized as a second array behind the value array), ``b'' is 2-D with variances, ``c'' is 1-D, and ``d'' is a scalar (0-D).
The three dimensions `x', `y', and `z' each come with an associated coordinate (green).
`y' has an additional auxiliary coordinate named ``labels'' (blue).
Finally, there is one scalar attribute named ``attr'' (red).
When viewed in a Notebook or a web browser, tooltips provide further information such as physical units and array shapes.
}
\end{figure}

At first the apparent complexity of such higher-level data structures compared to, e.g., a plain \numpy \texttt{ndarray}, may be overwhelming for new users.
Therefore, \scipp supports programmatic generation of a visualization of its data containers, e.g., for a dataset as shown in Fig.~\ref{fig:show}. 
This greatly simplifies writing of easy-to-read documentation and helps illustrating inside \jupyter Notebooks for teaching purposes.
Furthermore, it is convenient as a quick tool to inspect in-memory data in day-to-day use when working in a \jupyter Notebook.

Operations with \scipp's data structures frequently make use of \emph{slicing}.
Just like \numpy, a slice operation in \scipp returns a \emph{view}, i.e., no copy is made.
Similarly, accessing an entry of a dataset returns a view that is equivalent to a data array, without making a copy.
In both cases, views can be used to modify the content of the data container they are referencing.
However, views prevent any modification that would break invariants of the data container.
For example, a slice of a variable cannot be used to modify the unit of the variable, since this would require storing a different unit for different elements of the same variable.
Similarly, a view of a data array cannot be used to change shape or dimension labels, since this would break the invariant of the dataset guaranteeing that dimensions of all data in the set are aligned.

\begin{figure}
\begin{lstlisting}
delta = data['a']['x', 1:3]  -  scipp.mean(data['b'], 'z')
# see text   (1)  (2)      (4-8)     (4,5)      (1)   (3)
\end{lstlisting}
\caption{\label{fig:operation}
Computation with \scipp's \python interface combining item access, slicing, function application, and binary operators.
Based on the two inputs on the right-hand side the resulting \texttt{delta} is a data array with units, dimensions, and coordinates given by the result of the slicing and computation.
}
\end{figure}

Structuring information into the categories data, coordinates, and attributes and characterizing them with names, dimension names, and units is essentially a handle for controlling the behavior of operations.
As an example, consider the line of code in Fig.~\ref{fig:operation}, processing the dataset from Fig.~\ref{fig:show}.
A Python user who has previously worked with \numpy will quickly be able to grasp the meaning and intention of this code, without much explanation:
We select a data item from a dictionary-like object by its name, slice it along a certain dimension, and then subtract the mean along another dimension of a different data item.
We break down what is happening under the hood to highlight the hidden power here:

\begin{enumerate}
  \item Data items are selected by name (here: ``a'' and ``b'').
  \item Slice based on \emph{named} dimension (here: slice ``a'' along `x').
  \item Apply other operation based on named dimension (here: \texttt{mean} along `z').
  \item Propagate uncertainties (here: \texttt{mean} and \texttt{-} (subtraction)).
  \item Handle physical units (here: \texttt{mean} and \texttt{-} (subtraction)).
  \item Match coordinates (here:\texttt{-} (subtraction)).
  \item Broadcast into missing dimensions (here:\texttt{-} (subtraction)).
  \item Transpose matching dimensions (not required here).
\end{enumerate}
With the exception of (4) and (5) this is exactly what \xarray also supports, unless event data is involved.
For anyone familiar with \numpy, items (2), (3), (7), and (8) bring an immediate improvement since they avoid the need to keep track of and remember dimension order, to explicitly broadcast, and to explicitly transpose.
Altogether, \scipp's data structures enable a brief and intuitive syntax that nevertheless handles all required concepts under the hood.
With correctness of the programmatic operations built into the framework, users are free to focus on the \emph{intent} when writing code for their scientific purposes.

\subsection{Event data}\label{sec:sparse-data}

Processing neutron-scattering data requires dealing with \emph{event data}~\cite{PETERSON201524}.
Event data is a certain form of unaligned sparse data arising from recording the time-of-flight of every detected neutron alongside metadata such as a pixel identifier and an index of the proton pulse that produced the neutron.
The data stored per neutron is tiny, typically between 1 and 4 integers or floating-point values, i.e., between 4~Byte and 32~Byte, depending on whether single-precision or double-precision is used.
As an in-memory representation we commonly store a separate list of neutron events for every pixel.
Structurally, event data amounts to an array of random-length lists --- not to be confused with a sparse matrix or a sparse array which may represent a matrix with many ``zero'' entries in a packed format.
In other words, the latter is \emph{aligned} but sparse data, whereas event data is \emph{unaligned}.
This is essentially raw data for creating a histogram for every pixel, i.e., raw data for a 2-D (or higher) array.

\begin{figure}
  \includegraphics[width=0.45\textwidth]{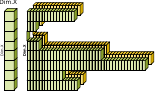}
\caption{\label{fig:show-sparse}One-dimensional data array with event data (yellow), the raw data for a two-dimensional histogram.
In addition to the coordinate for `x' (green), there is an event coordinate (green) matching the length of the data.
An event coordinate is always required if there is event data since by definition every data value is ``measured'' at a different coordinate value.}
\end{figure}

Due to the importance of event data in the field, \scipp is designed to support this particular form of sparse data as arrays with list-valued elements.
Figure~\ref{fig:show-sparse} is an example of a data array containing event lists.
Internally \scipp is using an array of vectors to store event data that supports:
\begin{itemize}
  \item 
    Element-wise operations between dense data and event data.
    The dense operand is broadcast into the internal ``dimension'' of the event lists, i.e., for a given event list every list entry is multiplied by the same dense value.
    Speaking in neutron terms, we can thus, e.g., add a pixel-dependent offset to the time-of-flight of all events.
  \item Slicing event data, but only along the usual dimensions.
    There is no sensible definition of index-based slicing the event lists themselves.
  \item Shape-changing operations such as concatenation of arrays in both dense dimensions and along the internal dimension of the event lists.
\end{itemize}

\section{Technical details and usage}\label{sec:details}
\subsection{Methodologies and distribution}\label{sec:methodologies}

\begin{figure}
  \centering
    \includegraphics[width=0.5\textwidth]{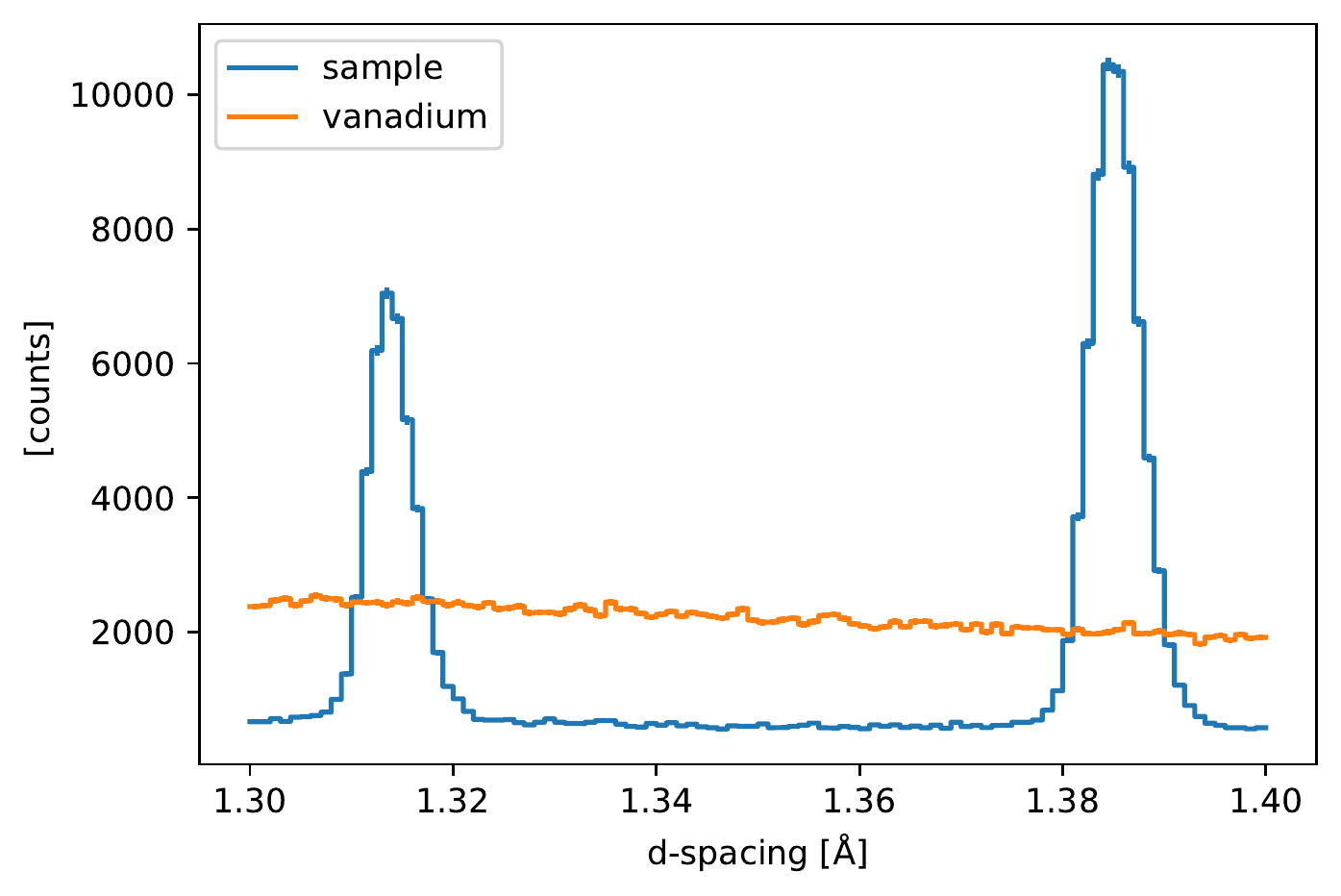}
  \caption{\label{fig:plot}
  Example of plotting with \scipp, demonstrating how the \dataarray and \dataset containers enable meaningful plots by default:
  (1) Axes are labeled and named based on coordinate values, coordinate names, and coordinate units.
  (2) Data units are used to name the data axis.
  (3) Multiple dataset entries yield plots with multiple named entities plotted (``sample'' and ``vanadium'' in this case).
  (4) Variances stored with data are used to plot standard deviations (error bars).
  (5) Coordinates can represent bin-edges, resulting in data plotted as histograms.
  Not shown in figure:
  If a dataset contains, e.g., 1-D and 2-D data, multiple subplots are generated.
  For 3-D (or higher) \scipp provides means to interactively slice and visualize data.
}
\end{figure}

\scipp is open source, licensed under the GPLv3, and is currently developed by the European Spallation Source (ESS).
It is hosted on GitHub (see \url{https://github.com/scipp/scipp}) and is developed using up-to-date practices such as code reviews, unit testing, and continuous integration.
The released version 0.3.1 is for experimental use and not production-ready yet.
We provide \conda packages for easy installation and integration in existing \python environments.

\scipp is implemented in \cpp.
This choice gives maximum freedom in terms of performance optimization and will allow for interfacing with existing \cpp applications like \mantid.
The use of the recent C++17 revision of the standard played a central part in keeping the library flexible, small, and maintainable --- despite its flexibility and number of features \scipp is currently just 10,000 lines of code of \cpp, not counting unit tests and benchmarks.
It is intended that most users would interface with \scipp using its \python bindings, which are implemented using pybind11~\cite{pybind11}.
On the \python side, \scipp is designed to play nicely with \numpy.
In particular, value and variance arrays in a variable can be initialized from \numpy arrays, and \numpy functions can be directly applied to those arrays.
This allows for integration with a large variety of other scientific software.

\jupyter Notebooks are rapidly becoming a go-to solution for scientific data handling, combining advantages from scripting interfaces with those of a graphical user-interface.
Therefore, in addition to visualizations as shown in Fig.~\ref{fig:show}, \scipp also provides table-displays for 1-D data and easy-to-use plotting functionality for any number of dimensions based on \matplotlib~\cite{Hunter:2007}.
Labeled dimensions, coordinates, physical units, and uncertainties stored in a data array or dataset alongside data allow for direct creation of meaningful plots without the need for additional user-provided information.
An example is given in Fig.~\ref{fig:plot}.
Most of \scipp's documentation (see \url{https://scipp.github.io}) is also written as \jupyter Notebooks and can thus be downloaded, interacted with, and experimented with to serve as a starting point for new users.
The documentation also includes a number of tutorials in the form of \jupyter Notebooks, such as the one displayed in the next section.

During development of \scipp we favor clean code over hypothetical performance gains.
When we \emph{do} implement concrete optimizations, we drive and \emph{justify} this using benchmarks modeling real workloads.

\subsection{Usage example}\label{sec:example}

\begin{figure}
\small
  \centering
  \begin{subfigure}{.50\textwidth}
Imports for \scipp and \numpy:
\begin{ipythonnb}
import numpy as np
import scipp as sc
from scipp.plot import plot
\end{ipythonnb}

Loading NeXus files:
\begin{ipythonnb}
tof_sam = sc.neutron.load('PG3_4844_event.nxs')
tof_van = sc.neutron.load('PG3_4866_event.nxs')
\end{ipythonnb}

Convert units to $d$-spacing (interplanar lattice spacing):
\begin{ipythonnb}
sam = sc.neutron.convert(tof_sam, 'tof', 'd-spacing')
van = sc.neutron.convert(tof_van, 'tof', 'd-spacing')
\end{ipythonnb}

Histogram event data:
\begin{ipythonnb}
bins = sc.Variable(dims=['d-spacing'],
                   unit=sc.units.angstrom,
                   values=np.arange(0.3, 2.0, 0.001))
h = sc.Dataset({'sample':sc.histogram(sam, bins),
                'vanadium':sc.histogram(van, bins)})
\end{ipythonnb}

Create plot of counts depending on $d$-spacing and spectrum:
\begin{ipythonnb}
plot(hist['sample']['d-spacing',100:400])
\end{ipythonnb}
\quad\includegraphics[width=0.9\textwidth]{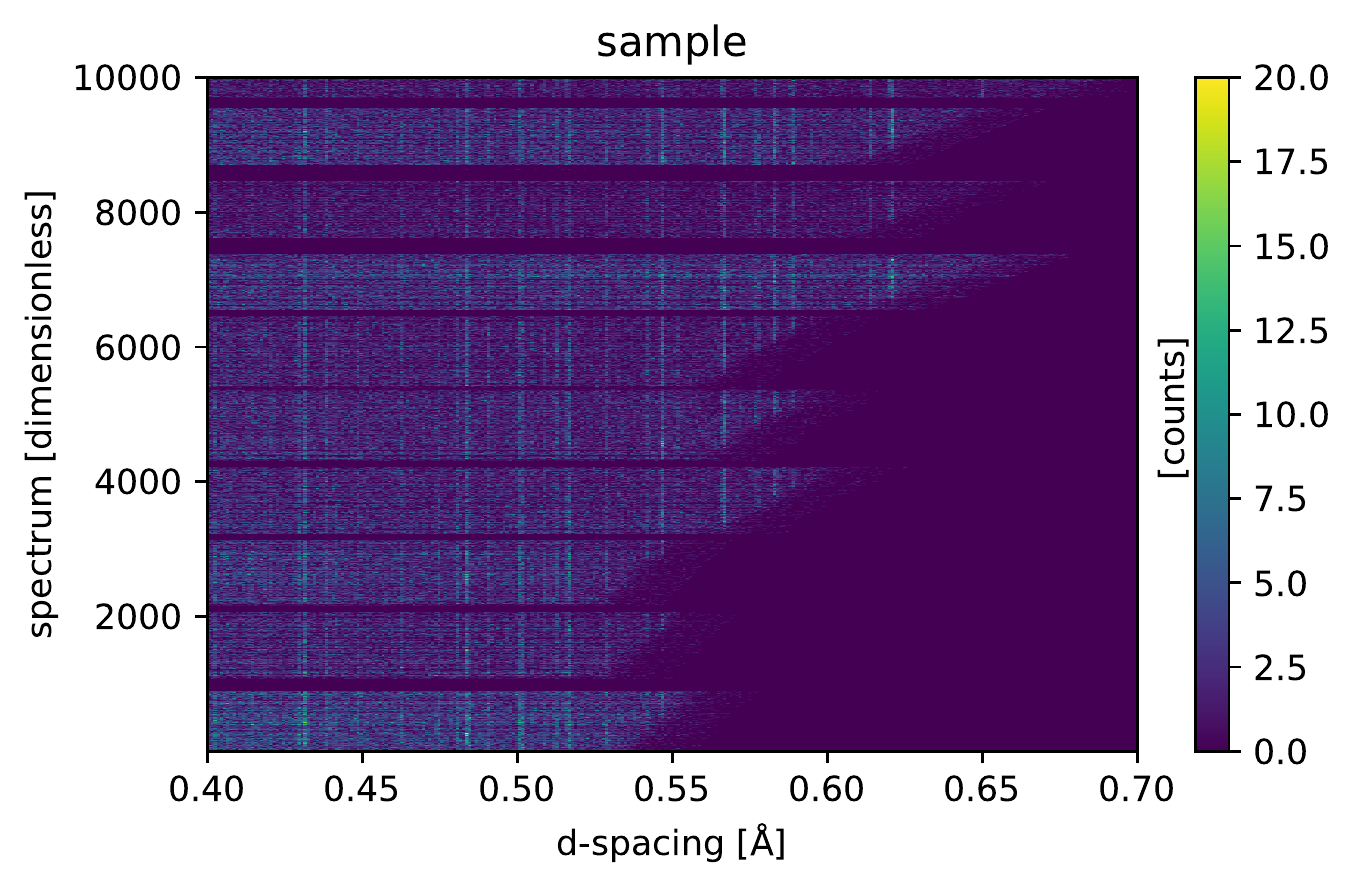}

Reduce data into a single spectrum using \texttt{scipp.sum}:
\begin{ipythonnb}
summed = sc.sum(hist, 'spectrum')
plot(summed)
\end{ipythonnb}
\quad\includegraphics[width=0.9\textwidth]{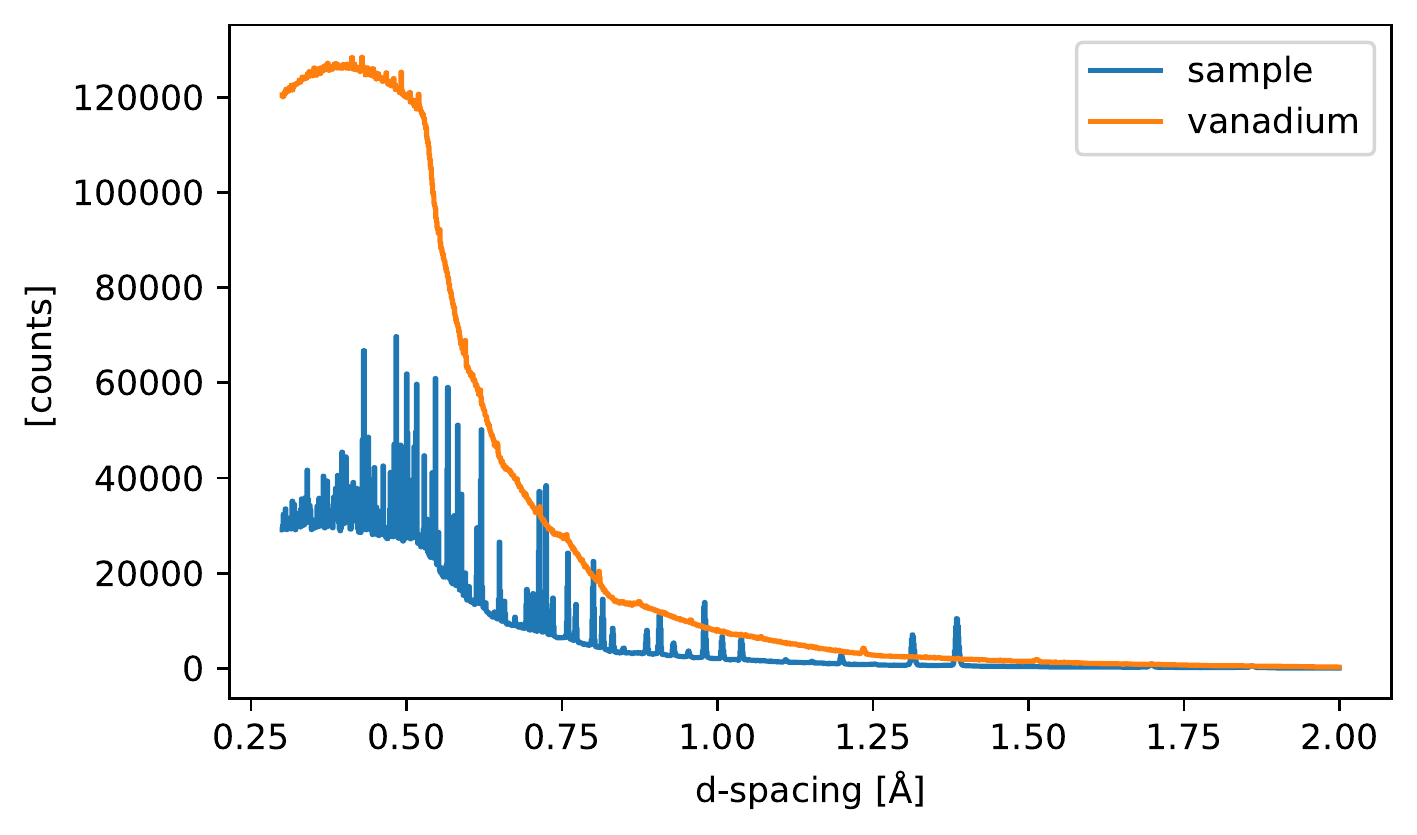}

\end{subfigure}%
\hfill
\begin{subfigure}{.48\textwidth}

Normalize sample data to vanadium:
\begin{ipythonnb}
normalized = summed['sample'] / summed['vanadium']
plot(normalized)
\end{ipythonnb}
\quad\includegraphics[width=0.9\textwidth]{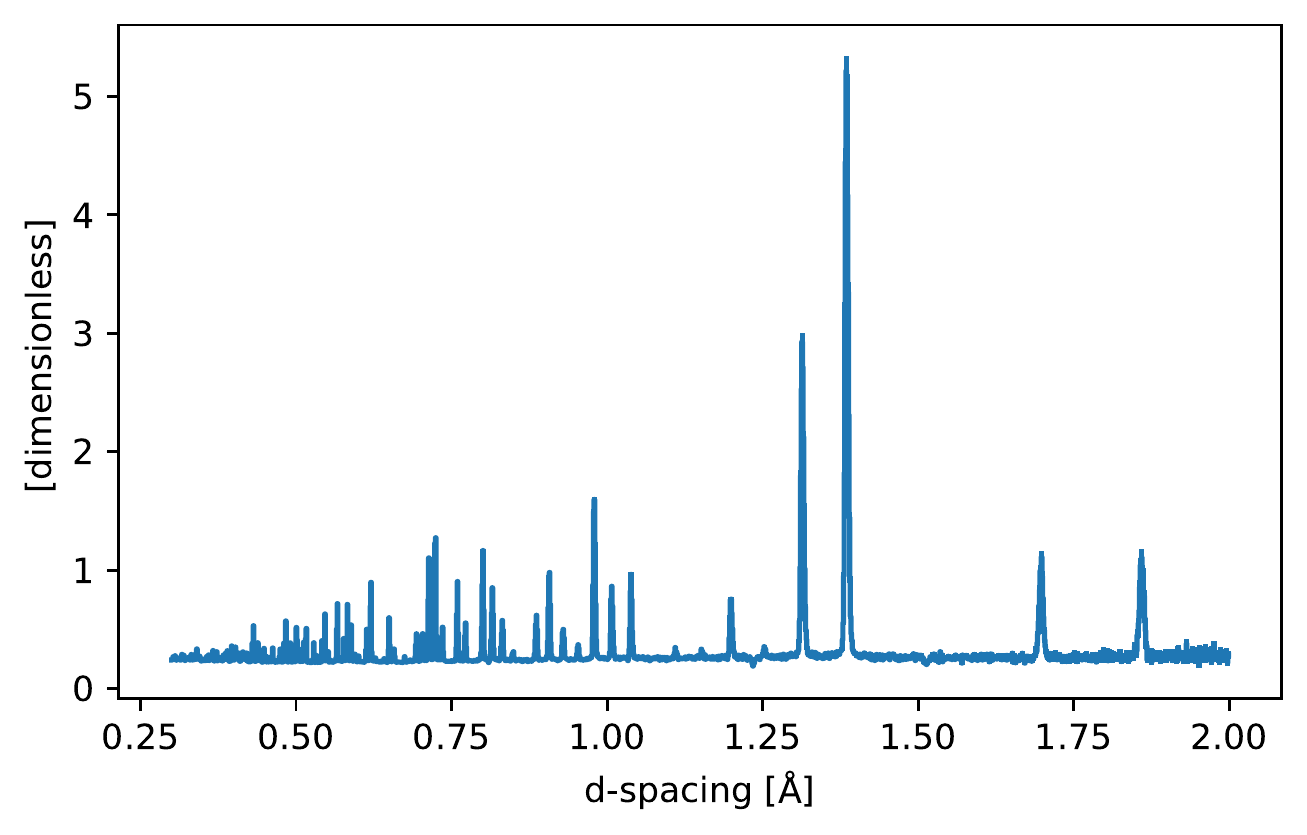}

Plots of quantities against time-of-flight or $d$-spacing and spectrum are often hard to read since ``spectrum'' is not a dimension with physical meaning but an instrument artifact.
\scipp's \texttt{groupby} provides a convenient tool for converting ``spectrum'' to something more useful.
We group and combine data based on scattering angle $\theta$ using \texttt{groupby}:
\begin{ipythonnb}
theta = sc.neutron.scattering_angle(tof_sam)
tof_sam.coords['theta'] = theta
theta_bins = sc.Variable(
    dims=['theta'], unit=sc.units.rad,
    values=np.linspace(0.5, 1.2, num=1000))
# Note: Use `sum` instead of `flatten` when working
# with dense (histogrammed) data
theta_tof_sample = sc.groupby(
    tof_sam,
    'theta',
    bins=theta_bins).flatten('spectrum')
\end{ipythonnb}

Create plot depending on time-of-flight and $\theta$, showing intensity peaks following a $\sin$, as expected from Bragg's law:
\begin{ipythonnb}
tof_bins = sc.Variable(
    dims=['tof'], unit=sc.units.us,
    values=np.linspace(10000, 16000, num=500))
plot(theta_tof_sample, bins={'tof':tof_bins})
\end{ipythonnb}
\quad\includegraphics[width=0.9\textwidth]{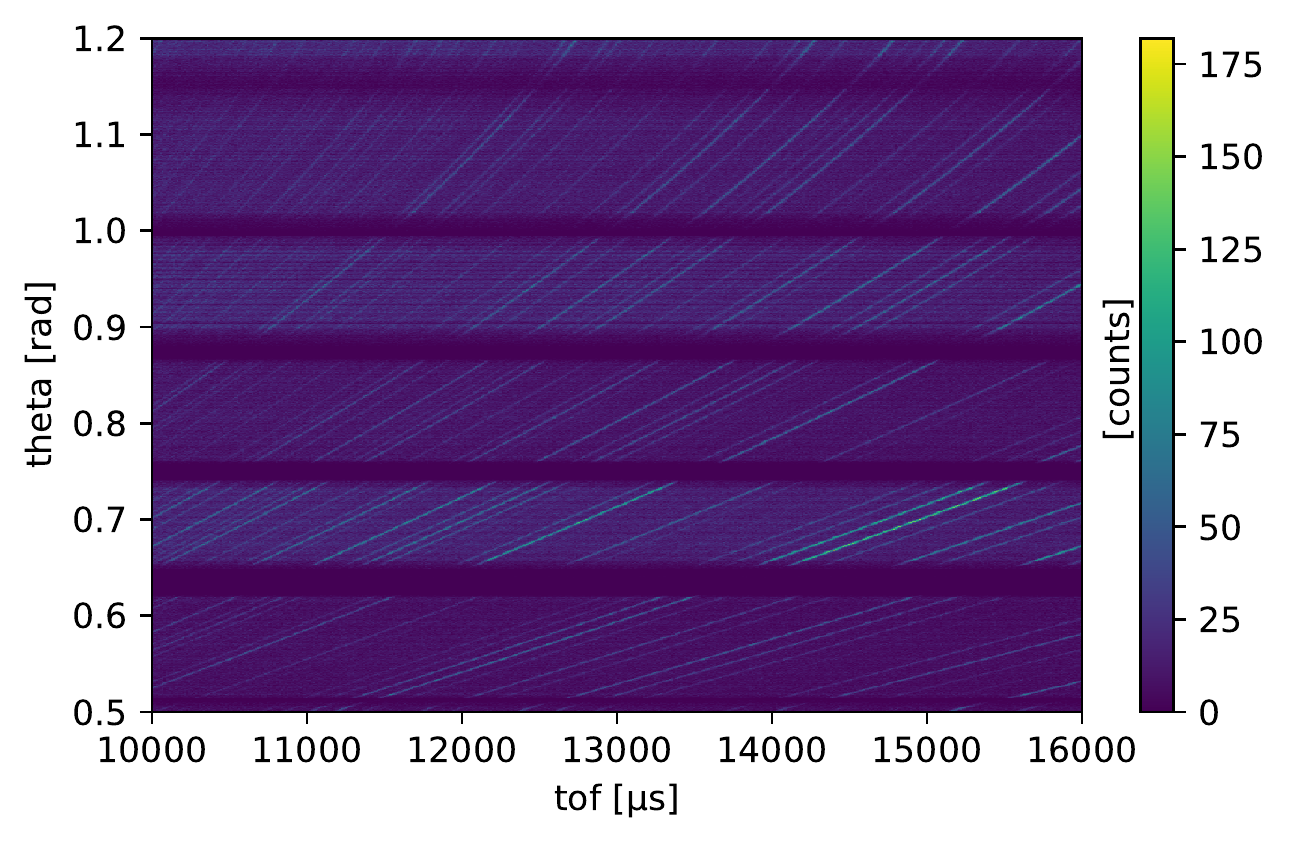}

  \end{subfigure}
  \caption{\label{fig:example}
Example of \scipp-usage in a \jupyter Notebook for a (simplified) data reduction workflow for time-of-flight powder-diffraction data in event mode.
See the electronic version of the article for high-resolution figures.
}
\end{figure}

To give a better understanding of the mechanics of \scipp and of its current capabilities, a concrete usage example is provided in Fig.~\ref{fig:example}.
The depicted workflow uses \scipp in a \jupyter Notebook to process neutron-scattering data.
To avoid an abrupt loss of encapsulated features and know-how when switching from using \mantid to using \scipp, conversion between the respective data containers is supported.
In this case this happens internally in \texttt{scipp.neutron.load}, which is actually relying on \mantid to load \nexus~\cite{Konnecke:po5029} files.
The resulting data array includes data and coordinates, as well as attributes holding the sample and run information, and a simplified instrument geometry.\footnote{In this case only pixel positions are required, so the instrument geometry in \scipp does not include pixel shapes or rotations.}

\subsection{Architecture}\label{sec:architecture}

\subsubsection{Type erasure}

\scipp's \variable must support a considerable number of types --- just our basic use cases include \texttt{double}, \texttt{float}, \texttt{int64\_t}, \texttt{int32\_t}, \texttt{std::string}, \texttt{bool}, \texttt{DataArray} for storing complex information such as neutron monitors, \texttt{Eigen::Vector3d} for positions in space, and event lists containing \texttt{double},  \texttt{float}, \texttt{int64\_t}, and \texttt{int32\_t}.
In \cpp this is obviously addressed using \emph{templates}.
However, since variables are combined in data arrays and datasets, a hypothetical templated \texttt{class Variable} would lead to a combinatoric explosion in the number of required template instantiations for the \texttt{DataArray} and \texttt{Dataset} classes.
Once we also consider that operations between data arrays with different type content are required, it becomes clear that this problem is intractable.\footnote{While in practice only a limited subset of type combinations will be used, \scipp is mainly used through its \python scripting interface, so we have to compile and build all possible type combinations into the library.}

To support storage of the various required types of data and combining arbitrary variables in data arrays or datasets, \scipp thus employs \emph{type erasure}~\cite{type-erasure}.
That is, \variable is not a templated class, i.e., there is no \texttt{Variable<double>} and \texttt{Variable<float>}.
The actual data type of the elements is not reflected in the class type but is erased by an internal mechanism based on \emph{concept-based polymorphism}~\cite{Pirkelbauer:2008:RCC:1363686.1363734}.
The result is a simple high-level \cpp interface that feels similar to an interface in \python.
This high-level interface can be used as long as the existing operations are sufficient as building blocks.
When this is not the case a crucial difficulty with type-erasure becomes apparent:
developers must either (a) choose the data type when implementing an operation and fail if the input does not have the expected type or (b) provide an implementation for all possible combinations of input types.
The former is generally too limiting and contravenes the flexibility gained from generic data structures and type erasure.
The next section describes how \scipp facilitates the latter.

\subsubsection{Generic transform algorithms}\label{sec:transform}

To leverage the power of type-erasure without excessive templating and specializations in implementations of operations,\footnote{Including first and foremost a large number of built-in operations.} \scipp employs a --- to our knowledge --- novel mechanism.
One approach would be to implement all operations as \texttt{virtual} methods of the underlying \texttt{VariableConcept} used for type erasure.
However, for operations with two (or more) inputs this would require double-dynamic (or multiple-dynamic) dispatch which is cumbersome to implement in \cpp, quickly leads to an intractable amount of code, and would make adding user-defined operations virtually infeasible.
Our solution works as follows:
\begin{itemize}
  \item Use \texttt{C++17}'s type-safe \texttt{union} type, \texttt{std::variant}, with \emph{alternative types} \texttt{VariableConceptT<T>} for a selected set of types \texttt{T}.
    In practice this includes, e.g., \texttt{double}, \texttt{float}, \texttt{int64\_t}, and \texttt{int32\_t}, i.e., the set of known fundamental types that make up the majority of data.
  \item Store the additional \emph{alternative type} \texttt{VariableConcept} to handle any other data types that are not built-in or known to the core library.
    This is a fallback and implies that data stored using this alternative cannot make use of the mechanism described next.
  \item Provide \texttt{transform} and \texttt{transform\_in\_place} algorithms (in the spirit of \texttt{std::transform}) that can \emph{apply arbitrary lambdas or functors} to a set of input variables.
    An example is given in Fig.~\ref{fig:transform}.
    Internally this is using a visitation approach as provided by \texttt{std::visit} to branch to the correct instantiation based on the alternative types of all the input's respective underlying variants.
    In practice we cannot use \texttt{std::visit} due to the large combinatoric factor when considering all possible types of all inputs.
    Instead, \texttt{transform} requires a list of supported type combinations of inputs.
    This is then used to generate all required branches.
\end{itemize}
\begin{figure}
  \begin{lstlisting}[language=C++]
Variable radius(const Variable &varX, const Variable &varY) {
  return transform<tuple<double, double>, tuple<double, float>,
                   tuple<float, double>, tuple<float, float>>(
      varX, varY, [](const auto x, const auto y) {
        return sqrt(x * x + y * y);
      });
}
\end{lstlisting}
\caption{\label{fig:transform}
Example of the internal \texttt{transform} function used to compute $\sqrt{a^2 + b^2}$.
The call to \texttt{transform} requires as explicit template arguments a list of all type combinations of the input arguments \texttt{varX} and \texttt{varY} that are to be supported.
In practice \scipp has pre-defined sets of commonly used combinations, so this does not need to be spelled out explicitly in all cases.
The third function argument of \texttt{transform} is a functor, in this case a \emph{lambda}.
\texttt{transform} iterates over the value arrays (and, if present, variance arrays) of the inputs, transposing and broadcasting if required, and applies the functor to all elements, including the case of elements with attached uncertainties.
The unit is also transformed using the functor.
Based one the input's dimensions, \texttt{transform} automatically creates an output variable with the correct dimensions.
}
\end{figure}
The crucial point here is the visitation approach, bypassing the \cpp restriction preventing virtual methods from being templated, while the fallback alternative in the variant keeps the flexibility of storing arbitrary data.
\texttt{transform} automatically handles units, propagation of uncertainties, dense and event data (including mixing the two), transposition, and broadcasting.
Any operation implemented based on \texttt{transform} will automatically leverage all this functionality, and implicitly benefit from using a single unified implementation that is well tested.
It is also used internally for implementing the majority of element-wise operations between variables.

\section{Status and evaluation}\label{sec:evaluation}
\subsection{Performance}\label{sec:performance}

At the time of writing \scipp has not been optimized for performance in its entirety.
In particular it is currently single-threaded.\footnote{Preliminary support for multi-threading based on Intel Threading Building Blocks has been added to \scipp since the original version of this manuscript, but a detailed discussion is beyond the scope of this article and will be included in a future publication.}
The current target application of \scipp is data-reduction for neutron data.
Ignoring I/O, the majority of these workloads is bound by memory-bandwidth when individual reduction steps are considered in isolation.
This implies two things:
(1) It is comparatively simple to optimize code such that it hits the bandwidth limit, as long as memory allocation is avoided.
The majority of \scipp's basic operations are optimized to this effect.\footnote{
  Since there is no multi-threading yet this is the \emph{single-thread (or single-core) bandwidth}, which is typically lower than the maximum bandwidth the whole CPU or multi-socket system can sustain.
That being said, a single thread in a system with $N$ cores can typically utilize significantly more than $1/N$ of the total bandwidth.
We can conjecture that when multi-threading is introduced the existing optimizations would still support bandwidth-bound operation, provided that the workload is large enough to support efficient multi-threading.}
(2) To overcome this limit we must either move to multi-node parallelization, employ cache-blocking techniques to make better use of the CPU caches,\footnote{dividing data into small chunks that fit into, e.g., L2 or L3 caches and applying a \emph{series} of operations before proceeding with the next chunk} or both.

Part of the motivation for work on \scipp is to open the opportunity for experimentation and investigation of aspect (2) in more detail.
Such investigations would be virtually impossible within the \mantid framework due to its size.
The mentioned approaches each come with advantages and disadvantages as well as limitations which require a detailed analysis and prototyping.
The decision is further complicated by the fact that hdf5-based I/O, which is the most common format in the community, has very limited support for multi-threading but could see potential speedups from multi-process parallelism.
Specifically for data-reduction of neutron-scattering data, part of the remit of the \scipp development is to attempt to push the envelope of what can be handled on a single-node system with multi-threading, despite ever-increasing data volumes.
Multi-process and multi-node parallelization always add complexity --- even with modern solutions such as dask~\cite{dask} --- and it is often beneficial to avoid such parallelization if feasible.

\subsection{Limitations and shortcomings}\label{sec:limitations}

Every choice has its downsides and also \scipp comes with a number of limitations that may or may not be significant, depending on the use case.

(a) \scipp does currently not support native I/O.
Instead, we leverage \mantid to load data from disk (or save it to disk), converting the loaded data into a data array or dataset after loading.
This certainly comes with a performance penalty and native \scipp I/O is likely to be implemented in the future, at least for file formats where performance is crucial.

(b) \scipp's current implementation of physical units is based on boost's~\cite{Schling:2011:BCL:2049814} units library.
The limitation here is that \texttt{boost::units} is a compile-time unit implementation, i.e., all possible (required) combinations of base units must be known at compile time and built into the library.
For specific scientific applications like data reduction of neutron-scattering data it should be possible to provide an exhaustive list of all required units.
In more general cases however \scipp will just produce an error message until the required unit is added, which requires recompilation of the library.
We hope that this is a temporary limitation until a suitable replacement of \texttt{boost::units} is available.

\subsection{Developing data reduction workflows}

One of the lessons learnt from Mantid is that algorithms are only useful if their contract is fulfilled, i.e., input data is expected to have certain dimensions, units, or metadata.
\scipp aims to overcome parts of --- but by no means all of --- these problems by relying on small and generic building blocks.
This reduces the number of preconditions and the contract is easier to fulfil.
By comparison, \mantid has a significant number of algorithms that do multiple things or attempt to keep their contract flexible by supporting a variety of inputs.
Over the years, additional special cases have been added to some of the existing algorithms.
As a consequence, contracts for a variety of algorithms have become large and unclear, and in many cases developers need to consult the \cpp implementation directly to understand limitations.
The difference between the two approaches is best illustrated by an example:

(a) \mantid's \texttt{DiffractionFocussing} is a specialized algorithm for powder reduction.
It supports two ways to provide grouping information, one of which involves loading a file.
Grouping is internally processed by specialized code.
Data in input workspaces may be event data or histogrammed data.
Processing can preserve events, or histogram the data on-the-fly.
If the input is histogrammed data the algorithm furthermore rebins data to a common grid.

(b) \scipp's approach is the opposite.
It turns out that the algorithm does not contain a single neutron-scattering specific step.
Instead we use several independent steps to achieve the equivalent:
(1) grouping information is loaded from files,
(2) grouping information is preprocessed using \texttt{groupby}\footnote{an implementation of the ``split-apply-combine'' approach known from \pandas and \xarray, see \url{https://scipp.github.io/user-guide/groupby.html}},
(3) data is reduced using generic \texttt{sum} (for histogrammed data) or \texttt{flatten} (for event data) operations, using \texttt{groupby} if multiple output spectra are required, and
(4) histogram explicitly if desired.

Each of the components in \scipp's way of implementing the workflow has a low complexity and a low number of conditions in their contract.
As a side benefit \scipp's approach also enables easier and more thorough unit testing and increases overall maintainability.

The part that \scipp cannot overcome is obviously that the overall complexity of the task it is used for will not simply disappear.
In the example above, at some point someone will need to write the code to execute steps (1-4).
\scipp's goal is to make that easier by providing small and generic but powerful building blocks.
This should give developers the opportunity to, e.g., choose to tailor workflows to a specific instrument without the need for complex algorithms supporting many cases, but at the same time avoiding large amounts of near-duplicate code.
It is too early to tell how well this will work out in practice, and the choice between specialized but complex higher-level algorithms and, e.g., instrument-specific workflows will need to be made on a case-by-case basis.

Lastly, as a side effect of being generic, a ``schema'' or ``standard'' needs to be followed when organizing data and metadata in a data array or dataset.
For example, neutron-specific algorithms like unit conversion will expect certain metadata in particular places, and will simply fail if this contract is violated.
In practice we essentially define this as part of our file loaders or converters from \mantid.
This is not very different from \mantid where despite fixed structures information may still be missing, e.g., missing instrument information such as the position of the sample will lead to a failure in \texttt{ConvertUnits}, i.e., \scipp's generality is not necessarily introducing a new problem.

\subsection{Related projects}\label{sec:related}

\subsubsection{Xarray}\label{sec:xarray}

As discussed in the introduction, \xarray has inspired \scipp heavily.
After several rounds of prototyping the similarities had become so big that we chose to adopt the same terminology, such as \dataset and \dataarray, to avoid cognitive overhead.
We should thus justify the choice of not using \xarray or not contributing changes to \xarray.
There is a number of aspects to consider:

(a) Physical units and propagation of uncertainties could be implemented with \xarray using NEP-18~\cite{NEP18}.
    This enhancement proposal for \numpy has been actively pushed by the \xarray developers for these and similar purposes.
    In the case of propagating uncertainties this comes with a potentially big caveat:
    Apart from simple functions like addition and subtraction, equations for uncertainty propagation of, e.g., multiplication, are non-trivial and can result in a 10x or more loss of performance when implemented naively based on \numpy array arithmetics, due to the need for allocating temporary arrays and streaming through data multiple times.
    That is, for decent performance more work is required --- to our understanding either (i) an essentially complete implementation of operations on arrays of data with uncertainties, i.e., exactly what we have done in \scipp, or (ii) an implementation of a new \texttt{dtype} in \numpy, storing data as arrays of value/variance pairs.\footnote{
      Option (a) hints at a potential opportunity to leverage features of \scipp with \xarray.
    The \texttt{\_\_array\_function\_\_} from NEP-18 could be implemented for \scipp's \variable, bringing support for physical units and uncertainties when used in conjunction with \xarray.}

(b) The type of sparse data required for handling event data is conceptually very different from dense \numpy arrays.
    Event data plays an absolutely central role for our current main objective, data reduction for neutron-scattering experiments.
    We do not just require \emph{support} for event data but also high \emph{performance}.
    Therefore we consider it essential that this component is written in \cpp as opposed to \python.\footnote{Event data as, e.g., an array of small \numpy arrays would not be adequate.}
    Since operations \emph{mixing} dense and event data are also required, implementing an event data array on its own is not sufficient, since this would lack (fast) operations with \numpy arrays which are used for the dense data in \xarray.
    That is, both have to be implemented, which already makes up a major part of code in \scipp.

(c) Histograms and more specifically bin edges require storing coordinates in a data array or dataset that exceed the data extent by 1.
    This is not supported directly in \xarray.
    There are certainly other options to do this, such as storing left and right bin edges as two coordinates or as a single coordinate storing a pair of the two, but this comes with other drawbacks and additional implementation effort.
    Bin-edge axes also allow for multi-dimensional histogramming, which will be part of a future version of \scipp.

Essentially we consider the list of additional requirements on top of what \xarray provides as too long and, in the case of event data as too fundamental.
Effecting and contributing such fundamental changes to an existing framework is a long process and likely not obtained within the available time frame.\footnote{before ESS operations start, i.e., a couple of years}
Furthermore, some of the requirements are unlikely to be obtainable within \xarray itself since they may be incompatible with the project's goals or philosophy.

\subsubsection{Xtensor}\label{sec:xtensor}

The \cpp library \xtensor~\cite{xtensor} provides data structures and operations similar to those in \numpy.
Its key feature is the use of \emph{expression templates} which allow for generation of efficient code for compound operations by, e.g., making good use of CPU caches and avoiding allocation of temporaries.
It could thus be considered as a candidate to replace the lower-level components of \scipp.
However, since \scipp relies on type-erasure to implement its higher-level concepts this is not possible in a convenient way --- we are not aware of a way to combine expression templates and type erasure.
A hypothetical use of \xarray in \scipp would thus have a small scope and would provide limited advantage.
The \texttt{transform} mechanism with arbitrary lambdas described in Sec.~\ref{sec:transform} provides similar benefits to expression templates and is successfully combined with type erasure.

\subsubsection{TensorFlow}\label{sec:tensorflow}

While conceptually serving a different purpose, we mention TensorFlow~\cite{tensorflow2015-whitepaper} as a related project.
TensorFlow supports dense, sparse, and ragged tensors.
Scipp provides no equivalent to TensorFlow's sparse tensors.
The TensorFlow \texttt{RaggedTensor} is equivalent to nested variable-length lists, corresponding to {\scipp}'s event data.

\subsubsection{Mantid}\label{sec:mantid}

As discussed throughout this article, \scipp is to a large extent based on similar requirements as \mantid.
\mantid is the current \emph{de facto} standard for data reduction at many neutron scattering facilities.
\mantid supports event data as well as dense data.
It is highly specialized and dedicated for handling neutron (and muon) data.
Multi-dimensional data is handled using \texttt{MDHistoWorkspace} and \texttt{MDEventWorkspace} but the support is less complete than for \mantid's central data container, \texttt{MatrixWorkspace}, which provides a common interface for a list of histograms or a list of event lists.
Workspaces provide uncertainty propagation for data values.
Physical units are available only for the time-of-flight-derived axes.
Data types and workspace contents are typically fixed and, e.g., multiple axes per dimension are not supported.
\mantid's \texttt{TableWorkspace} can hold any number of columns and is thus similar to a 1-D \scipp dataset, but lacks support for coordinates and is not readily interoperable with other workspace types due a very different interface.
Finally, \texttt{WorkspaceGroup} can be seen as an equivalent to \scipp's \dataset and allows for applying operations to multiple workspaces at once.
It is actually even more generic since unlike \dataset it does not enforce coordinate alignment.

The functionality \mantid supports significantly exceeds \scipp's current capabilities.
Functionality is accessible in \cpp and via \python bindings.
Most higher-level are encapsulated in so-called \emph{algorithms}.
A long history with explicit, detailed, and inflexible data structures has led to inconsistent, incomplete, and hard to use (\python) interfaces, as well as an inflation of the number of algorithms, making maintenance increasingly harder.
Several years of work on data handling for ESS --- a new facility which needs to commission more than ten new instruments over the coming years~\cite{ANDERSEN2020163402} --- has proven that necessary changes to \mantid are often time-consuming or out-right impossible.
Furthermore, long training times and the recurrent need to make deep changes in \mantid due to lacking features or data access in the \python interfaces made it increasingly risky to continue relying solely on \mantid during the ESS hot-commissioning phase.
On the plus side, when used via the graphical user interfaces for existing and well-supported instruments at existing facilities, \mantid is very powerful and full-featured.

As the \emph{de facto} standard, \mantid encapsulates more than a decade's worth of knowledge about handling neutron-scattering data.
The \scipp project aims to leverage this by using \mantid as a \python library for specialized tasks.
Current examples of this are loading \nexus files and sample-geometry-based corrections such as absorption corrections.
\scipp's goal is not to fully replace \mantid --- we rather foresee that the two projects will continue to live in parallel, exploiting synergies and, for some aspects, living in symbiosis.

\section{Conclusion and outlook}\label{sec:conclusion}

We have presented an early version of the \scipp library.
\scipp provides comparatively simple data structures alongside a set of generic operations.
With flexibility designed into the data structures, it can nevertheless be a powerful tool to tackle many tasks involving scientific data.
For example, \scipp appears to be capable of adequately representing and processing neutron-scattering data without the need for dedicated data structures, and with reduced need for dedicated algorithms.
%Despite its generality and simplicity, \scipp appears to be capable of adequately representing and processing neutron-scattering data.
%While being capable of adequately representing and processing neutron-scattering data, \scipp is simple and general enough to be usable elsewhere.
%TODO list features 1-D data 3-D or higher data, arbitrary types and axes, notebook integration, \ldots?
A number of \scipp's features discussed throughout this article already surpasses the current scope of \mantid.
What is lacking is, e.g., support for the multitude of specialized correction algorithms that \mantid provides, but users of \scipp can rely on manual or automatic conversion of data from \mantid to \scipp.
For example, we reuse \mantid algorithms that rely on advanced geometry operations by providing wrappers which setup the input workspaces and convert output workspaces back to \scipp.

While \scipp is and will be kept generic, development effort is currently focussed on concrete feature and performance requirements for data reduction at the ESS.
That is, while we are consolidating common basics, significant amounts of work will be going into event-data handling and performance improvements in operations with event data.
To avoid going off on a tangent, we are using real workflows --- initially data reduction for an ESS powder diffractometer with several million pixels and event rates exceeding $10^7~\mathrm{neutrons/s}$ --- to steer and drive development.
As discussed in Sec.~\ref{sec:performance}, a naive implementation typically ends up being bandwidth-bound.
When reaching that point we will evaluate the balance between I/O and computation for the aforementioned model workflow.
This will flow into a decision on which parallelization strategies to focus on.
Other future work will include direct support for loading and saving files as well as an improved interface for working with event data or other unaligned data.

The final publication is available at IOS Press through \url{http://dx.doi.org/10.3233/JNR-190131}.

%%%%%%%%%%% The bibliography starts:

%%  ios1.bst will be used to                               %%

% if your bibliography is in bibtex format, use those commands:
\bibliographystyle{ios1}           % Style BST file.
\bibliography{bibliography}        % Bibliography file (usually '*.bib')

% or include bibliography directly:
%\begin{thebibliography}{0}
%\bibitem{r1} F. Author, Information about cited object.
%
%\bibitem{r2} S. Author and T. Author, Information about cited object.
%\end{thebibliography}

\end{document}